\documentclass[12pt]{article}
\usepackage{graphicx}
\usepackage{hyperref}
\usepackage{amsmath}
\usepackage{amscd,amssymb}
\usepackage{xcolor}

\newcommand{\fA}{\mathfrak{A}}
\newcommand{\fB}{\mathfrak{B}}

\newcommand{\nB}{\textsf{B}}

\newcommand{\vn}{\overrightarrow{n}}
\newcommand{\vepsilon}{\overrightarrow{\epsilon}}

\def\mg{{\mathfrak g}}

\def\ml{{\mathfrak l}}
\def\ms{{\mathfrak s}}
\def\mg{{\mathfrak g}}

\def\mg{{\mathfrak g}}

\def\ml{{\mathfrak l}}
\def\mg{{\mathfrak g}}

\def\cR{{\mathcal R}}

\def\cN{{\mathcal N}}

\def\cC{{\mathcal C}}

\def\mg{{\mathfrak g}}

\def\ml{{\mathfrak l}}
\def\mg{{\mathfrak g}}

\newtheorem{remark}{Remark}[section]

\def\e{{\,\rm e}\,}

\newcommand{\beq}{\begin{eqnarray}}
\newcommand{\eeq}{\end{eqnarray}}
\numberwithin{equation}{section}

\begin{document}

\begin{center}
{\large\bf Partition Functions for Supersymmetric Gauge Theories on Spheres}
\end{center}

\vspace{0.1in}

\begin{center}
{\large
A. A. Bytsenko $^{(a)}$
\footnote{E-mail: aabyts@gmail.com},
M. Chaichian $^{(b)}$
\footnote{E-mail: masud.chaichian@helsinki.fi}
and A. E. Gon\c{c}alves $^{(a)}$
\footnote{E-mail: aedsongoncalves@gmail.com}}
\end{center}

\begin{center}
\vspace{2mm}
$^{(a)}$
{\it
Departamento de F\'{\i}sica, Universidade Estadual de
Londrina\\ Caixa Postal 6001,
Londrina-Paran\'a, Brazil}

\vspace{0.5cm}
$^{(b)}$
{\it
Department of Physics, University of Helsinki\\
P.O. Box 64, FI-00014 Helsinki, Finland}

\end{center}

\vspace{0.1in}

\begin{abstract}
In this paper we briefly review the main idea of the localization technique and its extension
suitable in supersymmetric gauge field theory. We analyze the partition function of the vector
multiplets with supercharges and its blocks on the even- and odd-dimensional spheres and squashed spheres.
We exploit so-called Fa\`a di Bruno's formula and show that multipartite partition functions
can be written in the form of expansion series of the Bell polynomials. Applying the
restricted specialization argument we show that $q$-infinite-product representation of partition
functions admits presentation in terms of the Patterson-Selberg (or the Ruelle-type) spectral
functions.
\end{abstract}

\vspace{0.1in}

\begin{flushleft}

 Partition functions in gauge supersymmetric
quantum field theory. The Bell polynomials and spectral functions of hyperbolic geometry.

\vspace{0.3in}
\end{flushleft}

\newpage

\tableofcontents

\section{Introduction}

One of the goals of this paper is to give the extension of the localization formula
to the path integral in the context of supersymmetric quantum field theory.
The localization technique can be applied to certain observables, such as partition
functions in supersymmetric gauge theory.

{\bf The plan of this paper is as follows.}
In Section \ref{Partition} we analyze partition functions for supersymmetric gauge theories on
spheres. The partition function of the vector multiplets with supercharges on the even- and
odd-dimensional spheres and squashed spheres is reviewed in Section \ref{Vector}. Partition functions
can be constructed by using the special functions $\Upsilon_r(x\vert\epsilon)$ and $S_r(x\vert\epsilon)$.
The main building block for these functions is the multiple Gamma function $\gamma_r(x\vert\epsilon)$.

Then in Section \ref{Bell} we use the so-called Fa\`a di Bruno's formula and
show that multipartite partition functions can be written in the form of expansion series by means
of the Bell polynomials.

Restricted specializations we analyze in Section \ref{Specializations}. We derive an infinite-product
formula for the partition function in the form of the Ruelle spectral functions, whose spectrum is encoded
in the Patterson-Selberg functions of the hyperbolic three-geometry.
In Section \ref{Symmetry} we describe the symmetry and modular properties in $q$-infinite-product
structure for an appropriate blocks of the partition functions.

Finally in Section \ref{Squashed} we describe the holomorphic block $B_\alpha$ associated with the
partition function on appropriate supersymmetric background. Then we compute the partition functions
in terms of the spectral Ruelle functions and briefly discuss some applications of these computations.

\section{Partition functions}
\label{Partition}

{\bf Motivations for localization technique in physics.}
In this paper we are interested in what can be said as a localization in (supersymmetric)
gauge field theory. Usually the localization technique applied to supersymmetric observables, such as partition
functions, supersymmetric Wilson loops, etc. It is important that the supersymmetric localization gives
opportunity to study the nonperturbative results for these class of observables. Also this is a powerful
tool for analysis of interacting quantum field theory. The localization answers can be given in terms of
complicated finite dimensional integrals. Therefore techniques to study of these integrals and the
relevant physical and mathematical information should be deduced and developed.

\subsection{The partition function of the vector multiplets}
\label{Vector}

Let us consider some examples of the supersymmetric gauge theories on spheres $S^d$. The attempt of
calculation of the partition function on $S^d$ in the case of a continuous range of $d$ has
been made in \cite{Minahan}. Useful analysis for localization of equivariant cohomology for compact and
non-compact group actions (the Duistermaat-Heckman formula; Harish-Chandra, Itzykson-Zuber integral formulas)
the reader can find in article \cite{BLW}. These results were generalized and extended to the squashed spheres.
Term {\it squashing} in other words means the homogeneous deformation (see also Section \ref{Squashed}).

The key principle that can be used for computation of the partition function is the localization argument.
The partition function of the vector multiplet with 4, 8 and 16 supercharges placed on the even- and
odd-dimensional spheres \cite{PZ} is given respectively by
\begin{eqnarray}
Z_{S^{2r}} & = &\int_{\mg\prime} da\prod_{w\in R_{Ad \mg}}\Upsilon_r(iw\cdot a\vert\epsilon)
e^{P_r(a)} + \cdots,
\label{P11}
\\
Z_{S^{2r-1}} & = &\int_{\mg\prime} da\prod_{w\in R_{Ad \mg}}S_r(iw\cdot a\vert\epsilon)
e^{P_r(a)} + \cdots.
\label{P22}
\end{eqnarray}
In Eqs. (\ref{P11}) and (\ref{P22}) the polynomial $P_r(a) = \sum_{j=1}^r\alpha_j{\rm Tr}(a^j)$ is coming from
the classical action of the theory, the integrals are taken over the Cartan subalgebra ${\mg\prime}$ of the
gauge Lie algebra $\mg$, the $w$ are weights of adjoint representation of $\mg$. The parameters $\alpha_j$
are associated to the Yang-Mills coupling, the Chern-Simons couplings and Fayet-Iliopoulos couplings \cite{Pestun}.
$\epsilon$-parameters are equivariant parameters of the $T^r\subset SO(2r)$ toric action on $S^{2r-1}$
(also $\epsilon_1, \ldots, \epsilon_r$ are the squashing parameters for $S^{2r}$).
The case of the odd and even dimensional spheres $S^{2r-1}$ and $S^{2r}$ leads to two type of special functions
called $S_r$ and $\Upsilon_r$ respectively, that can present the final result for the partition function.

\begin{remark}
The nature of the dots in Eqs. (\ref{P11}) and (\ref{P22}) for arbitrary $r$ is not well clear.
Indeed:
\begin{itemize}
\item For supersymmetric gauge theories on $S^2$ there is $\cN=2$ vector multiplet with 4 supercharges \cite{Benini,Doroud} and the
dots are nonperturbative contributions coming from localization {\it loci} with
nontrivial magnetic fluxes.

\item For $S^3$ there is $\cN=2$ vector multiplet with 4 supercharges (see Refs. \cite{Kapustin}
and \cite{Willett} for a review) and the dots are absent.

\item For $S^4$ there is $\cN=2$ vector multiplet with 8 supercharges \cite{Pestun}
and the dots correspond to the contributions of pointlike instantons. These
poles can be computed using  the Nekrasov instanton partition function
\cite{Nekrasov}.

\item For $S^5$ there is $\cN=1$ vector multiplet with 8 supercharges \cite{Kallen1,Kim} and there are no
systematic derivation and understanding of the dots (also for the case $S^7$).

\item For $S^6$ there is $\cN=2$ vector multiplet with 16 supercharges \cite{Minahan}, and the nature of the dots remains to be understood.

\item For $S^7$ there is $\cN=1$ vector multiplet with 16 supercharges \cite{Minahan}.
\end{itemize}
\end{remark}

The main building block for functions $S_r$ and $\Upsilon_r$ is the multiple Gamma function
$\gamma_r(x\vert \epsilon_1, \ldots, \epsilon_r)$, it is a function of a variable $x$ on
complex plane $\mathbb C$ and  $r$ complex parameters $\epsilon_1, \ldots, \epsilon_r$.

Function $\gamma_r(x\vert \epsilon_1, \ldots, \epsilon_r)$ can be defined as a zeta-regularized
product
\begin{equation}
\gamma_r(x\vert\epsilon) = \gamma_r(x\vert \epsilon_1, \ldots, \epsilon_r) = \prod_{n_1,\ldots, n_r=0}^\infty
(x+n_1\epsilon_1+\cdots + n_r\epsilon_r).
\label{g}
\end{equation}
In Eq. (\ref{g}) the parameters $\epsilon_j$ belong to an open half-plane of $\mathbb C$ bounded by a
real line passing trough the origin. Among other things the unrefined version of $\gamma_r$ is defined as
\begin{equation}
 \gamma_r(x) = \gamma_r(x\vert 1,\ldots, 1) =
 \prod_{k=0}^\infty (x+k)^{\frac{(k+1)(k+2)\cdots (k+r-1)}{(r-1)!}}.
\end{equation}
The localization on $S^{2r}$ based on the $\Upsilon_r$-function, which can be defined as
\begin{equation}
\Upsilon_r(x\vert\epsilon) = \Upsilon_r(x\vert \epsilon_1, \ldots,\epsilon_r)  =
\gamma_r(x\vert \epsilon_1, \ldots \epsilon_r)
\gamma_r\left(\sum_{j=1}^r \epsilon_j-x\vert \epsilon_1,\ldots, \epsilon_r\right)^{(-1)^r}.
\end{equation}

The unrefined version of $\Upsilon_r$ can be defined as follows
\begin{equation}
 \Upsilon_r(x) = \Upsilon_r(x\vert 1,\ldots,1) = \prod_{k\in {\mathbb Z}}
 (k+x)^{sgn(k+1)\frac{(k+1)(k+2)\cdots (k+r-1)}{(r-1)!}}.
\end{equation}
The $S_{r}$-function, as defined below (see Ref. \cite{Narukawa} for details),
plays an important role upon localization on $S^{2r-1}$:
\begin{equation}
S_r(x\vert\epsilon) = S_r(x\vert \epsilon_1,\ldots, \epsilon_r) =\gamma_r(x\vert \epsilon_1,\ldots, \epsilon_r)
 \gamma_r\left(\sum_{j=1}^r\epsilon_j-x\vert \epsilon_1,\ldots,\epsilon_r\right)^{(-1)^{r-1}}
\end{equation}

\subsection{Infinite-product representation for \texorpdfstring{$\Upsilon_r(x\vert \epsilon)$}{Upsilon\_r(x|epsilon)} and \texorpdfstring{$S_r(x\vert\epsilon)$}{S\_r(x|epsilon)}}
\label{Product}

By definition the infinite-product of the $\Upsilon_r(x\vert \epsilon)$ has the form \cite{PZ}:
\begin{equation}
 \Upsilon_r(x\vert \epsilon) \stackrel{\rm{reg}}{=} \prod_{n_1=0,\ldots,n_r=0}^\infty
 \left(x+\sum_{s=1}^rn_s\epsilon_s\right)\left(\epsilon-x+\sum_{s=1}^rn_s\epsilon_s\right)^{(-1)^r}\!\!\!\!\!.
 \label{Y}
\end{equation}
In Eq. (\ref{Y}) $\stackrel{\rm reg}{=}$ denotes zeta-function regularization.
Also the infinite-product definition for the $S_r(x\vert\epsilon)$ is \cite{PZ}:
\begin{equation}
S_r(x\vert \epsilon) \stackrel{\rm{reg}}{=} \prod_{n_1=0,\ldots,n_r=0}^\infty
 \left(x+\sum_{s=1}^rn_s\epsilon_s\right)\left(\epsilon-x+\sum_{s=1}^rn_s\epsilon_s\right)^{(-1)^{r-1}}\!\!\!\!\!.
 \label{S}
\end{equation}

$\Upsilon_{r}$-function (\ref{Y}) and $S_{r}$-function (\ref{S}) can be represented in the following forms

\begin{eqnarray}
 H_r(x\vert\epsilon) & = & \fA_r(x\vert \epsilon)\cdot\fB_r(x\vert \epsilon)^{(-1)^r},
 \\
 G_r(x\vert\epsilon) & = & \fA_r(x\vert \epsilon)\cdot\fB_r(x\vert \epsilon)^{(-1)^{r-1}},
 \end{eqnarray}
 where
\begin{eqnarray}
\fA_r(x\vert \epsilon) & = &
\prod_{n_1=0,\ldots,n_r=0}^\infty
 \left(1-e^{2\pi i x}e^{2\pi\vn \vepsilon}\right),
 \label{A}
 \\
\fB_r(x\vert \epsilon) & = &
\prod_{n_1=0,\ldots,n_r=0}^\infty
 \left(1-e^{2\pi i(\sum_{j=1}^r\epsilon_j-x)}e^{2\pi i\overrightarrow{n}\overrightarrow{\epsilon}}
 \right).
 \label{B}
\end{eqnarray}

\begin{remark}
On the time being our discussions can be summarized as follows:

$\Upsilon_r$-functions form a hierarchy with respect to a shift of $x$ by one of $\epsilon$-parameters
\[
\Upsilon_r(x+\epsilon_j\vert \epsilon_1,\ldots, \epsilon_j,\ldots,\epsilon_r)
= \Upsilon_{r-1}^{-1}(x\vert \epsilon_1,\ldots, \epsilon_{j-1},\epsilon_{j+1},\ldots \epsilon_r)
\Upsilon_r(x\vert\epsilon_1,\ldots, \epsilon_j,\ldots,\epsilon_r).
\]

$S_r$-functions make up a hierarchy with respect to a shift of $x$ by one of the $\epsilon$-parameters
\[
S_r(x+\epsilon_j\vert \epsilon_1,\ldots, \epsilon_j,\ldots,\epsilon_r) =
S_{r-1}^{-1}(x\vert \epsilon_1,\ldots, \epsilon_{j-1},\epsilon_{j+1},\ldots, \epsilon_r)
S_r(x\vert \epsilon_1,\ldots, \epsilon_j,\ldots, \epsilon_r).
\]

Expression for $\fB_r(x\vert\epsilon)$ can be obtained from $\fA_r(x\vert\epsilon)$
by shifting of $x$ by $\epsilon$-parameters: $\fB_r(x\vert\epsilon) =
\fA_r(x\mapsto \sum_{j=1}^r\epsilon_j-x\vert\epsilon)$.

\begin{eqnarray}
 H_{r(odd\,\,r)}(x\vert\epsilon) & = & \frac{\fA_r(x\vert\epsilon)}{\fB_r(x\vert\epsilon)},\,\,\,\,\,
\,\, H_{r(even\,\, r)}(x\vert\epsilon) = \fA_r(x\vert\epsilon)\cdot\fB_r(x\vert\epsilon),
 \\
G_{r(even\,\, r)}(x\vert\epsilon) & = & \frac{\fA_r(x\vert\epsilon)}{\fB_r(x\vert\epsilon)},\,\,\,\,\,\,\,
G_{r(odd\,\, r)}(x\vert\epsilon)  =  \fA_r(x\vert\epsilon)\cdot\fB_r(x\vert\epsilon).
\end{eqnarray}
\end{remark}

\section{The Bell polynomials}
\label{Bell}

Let us consider for any ordered $r$-tuple of nonnegative integers (not all zeros)
$\epsilon =(\epsilon_1, \epsilon_2, \ldots ,\epsilon_r)$ the (multi)partitions, i.e. distinct
representations of $(\epsilon_1, \epsilon_2, \ldots ,\epsilon_r)$ as sums of multipartite numbers.
Because of Eq. (\ref{A}) we have
\begin{eqnarray}
{\rm log}\, {\fA}_r(x\vert \epsilon) & = & \sum_{n_1=0, \ldots, n_r=0}^\infty
{\rm log}(1- e^{2\pi ix+2\pi i\vn\vepsilon})
\nonumber \\
& = & - \sum_{n_1=0, \ldots, n_r=0}^\infty \sum_{m=1}^\infty \frac{e^{2\pi ixm}}{m}e^{2\pi i n_1\epsilon_1m}\cdots
e^{2\pi i n_r\epsilon_rm}
\nonumber \\
& = & -\sum_{m=1}^\infty\frac{e^{2\pi ixm}}{m}(1-e^{2\pi i\epsilon_1m})^{-1}\cdots
(1-e^{2\pi i\epsilon_rm})^{-1}
\nonumber \\
& = & -\sum_{m=1}^\infty \frac{e^{2\pi ixm}}{m}
\prod_{j= 1}^r (1-e^{2\pi i \epsilon_jm})^{-1}.
\end{eqnarray}

Let $\psi(m\vert r):= \prod_{j=1}^r(1-e^{2\pi i \epsilon_jm})^{-1}$, then
\begin{eqnarray}
\fA_r(x\vert \epsilon) & = & \exp\left( -\sum_{m=1}^\infty \frac{e^{2\pi ixm}}{m}\psi(m\vert r)\right),
\\
\fB_r(x\vert \epsilon) & = & \exp\left( -\sum_{m=1}^\infty \frac{e^{2\pi i(\sum_{j=1}^r \epsilon_j-x)m}}{m}
\psi(m\vert r)\right).
\end{eqnarray}

Let us introduce the following series expansions:
\begin{eqnarray}
\fA_r(x\vert \epsilon) & = & 1+\sum_{m=1}^\infty \cC_me^{2\pi ixm},
\\
\fB_r(x\vert \epsilon) & = & 1+\sum_{m=1}^\infty \cC_me^{2\pi i(\sum_{j=1}^r\epsilon_j-x)m}.
\end{eqnarray}
The {\it Bell polynomials} are very useful in many problems in combinatorics, specially in view of extensive tables
of Bell polynomials. We would like to note their application for the multipartite partition problem \cite{Andrews1}.
In particular the Bell polynomials technique can be used for calculation of $\cC_m$.

The Bell polynomials, first extensively studied by E. T. Bell \cite{Bell}, arise in the task of taking
the $n$-th derivative of a composite function $h(t)=f(g(t))$. We can find a formula for the $n$-th derivative of
$h(t)$. Indeed, let us denote $d^nh/dt^n = h_n$, $d^nf/dg^n = f_n$ and $d^ng/dt^n=g_n$, then we have $h_1=f_1$,
$h_2=f_1g_2 +f_2g_1^2$, $h_3=f_1g_3 + 3f_2g_2g_1+ f_3g_1^3, \ldots$. By mathematical induction we find
$h_n = f_1\alpha_{n1}(g_1, \ldots, g_n)+ f_2\alpha_{n2}(g_1, \ldots, g_n)+ \cdots + f_n\alpha_{nn}(g_1, \ldots, g_n)$,
where $\alpha_{nj}(g_1,\ldots,g_n)$ is a homogeneous polynomial of degree $j$ in $g_1,\ldots, g_n$.
As a result, the study of $h_n$ may be reduced to the study of the Bell polynomials: $Y_n(g_1,g_2,\ldots, g_n)
= \alpha_{n1}(g_1,\ldots, g_n)+ \alpha_{n2}(g_1,\ldots, g_n)+\cdots +\alpha_{nn}(g_1,\ldots, g_n)$. We should stress
that $Y_n$ is a polynomial in $n$ variables and the fact that $g_j$ was originally an $j$-th derivative is
not necessary in the consideration.

Useful expressions for the recurrence relations of the Bell polynomials
$Y_{n}(g_1, g_2, \ldots , g_{n})$
and it generating function ${\nB}(z)$ have the forms \cite{Andrews1}:
\begin{eqnarray}
&& Y_{n+1}(g_1, g_2, \ldots , g_{k+1}) = \sum_{k=0}^n  \begin{pmatrix} n\cr k\end{pmatrix}
Y_{n-k}(g_1, g_2, \ldots , g_{n-k})g_{k+1},
\label{BP}
\\
&&
\nB(z) = \sum_{n=0}^\infty Y_n z^n/n! \Longrightarrow
{\rm log}\,{\nB}(z)= \sum_{n=1}^\infty g_n z^n/n!
\label{FB}
\end{eqnarray}
In order to verify the last formula (\ref{FB}) we need to differentiate it with respect to $z$ and observe
that a comparison of the coefficients of ${z}^n$ in the resulting equation produces an
identity equivalent to (\ref{BP}).

Recall that a partition of a positive integer n is a finite nonincreasing
sequence of positive integers $k_1, k_2,\ldots, k_r$ such that $\sum_{j=1}^rk_j = n$. The $k_j$ are
called the parts of the partition. The partition $(k_1, k_2,\ldots, k_r)$ will be denoted by $\bf k$, and we shall
write ${\bf k}\vdash n$ to denote {\it {\bf k} is a partition of n}.

From Eq. (\ref{BP}) one can obtain the following explicit formula for the
Bell polynomials (it is known as the Fa\'{a} di Bruno's formula)
\begin{equation}
Y_{n}(g_1, g_2, \ldots , g_{n}) = \sum_{{\bf k}\,\vdash\, n}\frac{n!}{k_1!\cdots k_n!}
\prod_{j=1}^n\left(\frac{g_j}{j!}\right)^{k_j}\!\!.
\end{equation}

Finally,  the following result holds (see also Ref. \cite{Andrews1}, Theorem
12.3)
\begin{equation}
\cC_m   =  \frac{1}{m!}Y_m \left( 0!\psi(1\vert r),\,\, 1!\psi(2\vert r)\,\,,
\ldots , \,\,(m-1)!\psi(m\vert r)\right),
\label{C}
\end{equation}
and
\begin{eqnarray}
\!\!\!\!\!\!\!\!\!\!\!\!
\fA_r(x\vert \epsilon) & = & 1+\sum_{m=1}^\infty \cC_me^{2\pi ixm}
\nonumber \\
\!\!\!\!\!\!\!\!\!\!\!\!
& =& 1+ \sum_{m=1}^\infty \frac{1}{m!}Y_m \left( 0!\psi(1\vert r),\,\, 1!\psi(2\vert r),\ldots , \,\,
(m-1)!\psi(m\vert r)\right)e^{2\pi ixm},
\\
\!\!\!\!\!\!\!\!\!\!\!\!
\fB_r(x\vert \epsilon) & = & 1+\sum_{m=1}^\infty \cC_me^{2\pi i(\sum_{j=1}^r\epsilon_j-x)m}
\nonumber \\
\!\!\!\!\!\!\!\!\!\!\!\!
& =& 1+ \sum_{m=1}^\infty \frac{1}{m!}Y_m \left( 0!\psi(1\vert r),\,\, 1!\psi(2\vert r),\ldots , \,\,
(m-1)!\psi(m\vert r)\right)
\nonumber \\
& \times & \!\!\! e^{2\pi i(\sum_{j=1}^r\epsilon_j-x)m}.
\end{eqnarray}
As an example, let us calculate $\cC_2$ coefficient:
\begin{eqnarray}
2\cC_2 & = & Y_2(\psi(1\vert r), \psi(2\vert r))
= \psi(1\vert r)^2 + \psi(2\vert r)
= \prod_{j=1}^r(1-e^{2\pi i \epsilon_j})^{-2}
\nonumber \\
& + & \prod_{j=1}^r(1-e^{4\pi i \epsilon_j})^{-1}.
\end{eqnarray}

\section{Restricted specializations}
\label{Specializations}

For some specializations, when $\{\epsilon_1, \epsilon_2, \ldots,\epsilon_r\} =
\underbrace{\{i\vartheta,i\zeta_2,i\zeta_3, ...,i\zeta_r\}}_r$ and $q :=e^{2\pi i\vartheta}$ we get
\begin{eqnarray}
\fA_r(x\vert \epsilon) & = &
\prod_{n_1=0,\ldots,n_r=0}^\infty
 \left(1-e^{2\pi i x}q^{n_1+n_2\zeta_2/\vartheta+\cdots + n_r\zeta_r/\vartheta}\right),
 \label{Aq}
 \\
\fB_r(x\vert \epsilon) & = & \fA_r(x\longmapsto i\vartheta+i\sum_{j=2}^r\zeta_j -x \vert  i\vartheta,i\zeta_2,\ldots, i\zeta_r).
\label{Bq}
\end{eqnarray}

{\bf Spectral functions of hyperbolic three-geometry.}
Let us begin by explaining some results on the Patterson-Selberg (the Ruelle type) spectral functions.
For details we refer the reader to \cite{BB,BCST} where spectral
functions of hyperbolic three-geometry were considered in connection with
three-dimensional Euclidean black holes, pure supergravity, and string amplitudes.

Let ${\Gamma}^\gamma \in G=SL(2, {\mathbb C})$ be the discrete group
\footnote{
From the point of view of the applications, homologies associated with algebras
$\mg = {\ms}{\ml}(N;{\mathbb C})$ important since they constitute the thechnical basis of the
proof of the combinatorial identities of Euler-Gauss-Jacobi-MacDonald \cite{Fuks}.
}
${\Gamma}^\gamma$ is defined by
\begin{eqnarray}
{\Gamma}^\gamma & = & \{{\rm diag}(e^{2n\pi ({\rm Im}\,\vartheta + i{\rm
Re}\,\vartheta)},\,\,  e^{-2n\pi ({\rm Im}\,\vartheta + i{\rm Re}\,\vartheta)}):
n\in {\mathbb Z}\} = \{{\gamma}^n:\, n\in {\mathbb Z}\}\,,
\nonumber \\
{\gamma} & = & {\rm diag}(e^{2\pi ({\rm Im}\,\vartheta + i{\rm
Re}\,\vartheta)},\,\,  e^{-2\pi ({\rm Im}\,\vartheta + i{\rm Re}\,\vartheta)})\,.
\label{group}
\end{eqnarray}
We may construct a zeta function of Selberg-type for the group
${\Gamma}^\gamma \equiv {\Gamma}_{(\alpha, \beta)}^\gamma$ generated by a
single hyperbolic element of the form ${\gamma_{(\alpha, \beta)}} = {\rm diag}(e^z, e^{-z})$,
where $z= \alpha +i\beta$ for $\alpha, \beta >0$. Actually $\alpha = 2\pi {\rm Im}\,\vartheta$
and $\beta = 2\pi {\rm Re}\,\vartheta$.
The Patterson-Selberg spectral function $Z_{{\Gamma}^\gamma} (s)$ for ${\rm Re}\, s> 0$ can be attached
to $H^3/{\Gamma}^\gamma$ as follows:
\begin{equation}
Z_{{\Gamma}^\gamma}(s) := \prod_{k_1,k_2\geq
0}[1-(e^{i\beta})^{k_1}(e^{-i\beta})^{k_2}e^{-(k_1+k_2+s)\alpha}].
\label{zeta00}
\end{equation}

The zeros of $Z_{\Gamma^\gamma} (s)$ are precisely the set of complex numbers
$
\zeta_{n,k_{1},k_{2}} = -\left(k_{1}+k_{2}\right)+ i\left(k_{1}-
k_{2}\right) \beta/\alpha + 2\pi i n/\alpha,$ with $n \in {\mathbb Z}$.
The magnitude of the zeta-function is bounded
for both ${\rm Re}\,s\geq 0$ and ${\rm Re}\,s\leq 0$, and its growth can be estimated as
\begin{equation}
\big|Z_{\Gamma^\gamma}(s) \big| \leq \Big(\,
\prod_{k_1+k_2\leq
|s|}\,  \e^{|s|\, \ell}\, \Big)\,
\Big(\, \prod_{k_1+k_2\geq
|s|}\, \big(1- \e^{(|s|-k_1-k_2)\, \ell} \big)\,\Big)
\leq C_1\,\e^{C_2\, |s|^3}
\label{estimate}
\end{equation}
for suitable constants $\ell,C_1, C_2$. The first product on the right-hand side of
(\ref{estimate}) gives the exponential growth, while the second product is bounded.
The spectral function $Z_{\Gamma^\gamma} (s)$ is an entire function of
order three and of finite type which can be written as a Hadamard product~\cite{BCST}
\begin{equation}
Z_{\Gamma^\gamma}(s) =
\e^{Q(s)} \
\prod_{\zeta \in {\Sigma}}\,
\Big(\, 1-\frac{s}{\zeta}\, \Big)\,
\exp \Big(\,
\frac{s}{\zeta} + \frac{s^2}{2\zeta^2} +
\frac{s^3}{3\zeta^3}\, \Big)\ ,
\label{Hadamard}
\end{equation}
where $\Sigma$ is the set of zeroes $\zeta := \zeta_{n,k_{1},k_{2}}$
and $Q(s)$ is a polynomial of degree at most three. (The product formula for entire
function (\ref{Hadamard}) is known as Weierstrass formula  (1876).)

For the next step let us introduce the Ruelle spectral function ${\mathcal R}(s)$ associated with
hyperbolic three-geometry \cite{BB,BCST}. The function ${\mathcal R}(s)$ is an
alternating product of more complicate factors, each of which is so-called Patterson-Selberg
zeta-functions $Z_{\Gamma^\gamma}$.
Functions ${\mathcal R}(s)$ can be continued meromorphically to
the entire complex plane $\mathbb C$, poles of ${\mathcal R}(s)$ correspond to zeros of
$Z_{\Gamma^\gamma}(s)$.
\begin{eqnarray}
\prod_{n=\ell}^{\infty}(1- q^{an+\varepsilon})
& = & \prod_{p=0, 1}Z_{\Gamma^\gamma}(\underbrace{(a\ell+\varepsilon)(1-i\varrho(\vartheta))
+ 1 -a}_s + a(1 + i\varrho(\vartheta)p)^{(-1)^p}
\nonumber \\
& = &
\cR(s = (a\ell + \varepsilon)(1-i\varrho(\vartheta)) + 1-a),
\label{R1}
\end{eqnarray}
being $\varrho(\vartheta) = {\rm Re}\,\vartheta/{\rm Im}\,\vartheta$,
$\sigma(\vartheta) = (2\,{\rm Im}\,\vartheta)^{-1}$,
$a$ is a real number, $\varepsilon, b\in {\mathbb C}$, $\ell \in {\mathbb Z}_+$.

Later on we will use the following notation: $(a)_n = a(a+1)\cdots (a+n-1)$ and
\begin{eqnarray}
(a; q)_n &=& \frac{(a;q)_\infty}{(aq^n; q)_\infty} = \prod_{m=0}^\infty [(1-aq^m)/(1-aq^{m+n})]
= (1-a)(1-aq)\cdots (1-aq^{n-1}),
\nonumber \\
&{}& n=1,2,\ldots,
\label{aq}
\end{eqnarray}
$n$ is a non-negative integer. The shifted $q$-factorial for $n=0$ is $(a; q)_n = 1$.
\footnote{
The following series expansion holds \cite{Andrews}:
$
(q;q)_\infty^{-a_n} = \prod_{n=0}^\infty(1-q^{n+1})^{-a_n} = 1+ \sum_{n=1}^\infty b_n q^n,
$
where $a_n$ and $b_n$ are integers. Then $nb_n  =  \sum_{j=1}^nA_jb_{n-j}$,\, where $A_j = \sum_{d\vert j}da_d$.
If either sequence $a_n$ or $b_n$ is given the other $A_j$ is uniquely determined.
}

Setting $e^{2\pi ix} q^{n_{1}+n_2\zeta_2/\vartheta + \cdots + n_r\zeta_r/\vartheta} = q^{x/\vartheta}\varOmega_{r-1}q^{n_1}$
with $\varOmega_{r-1}=
q^{n_2\zeta_2/\vartheta +\cdots +n_r\zeta_r/\vartheta},$ we get
\begin{eqnarray}
\!\!\!\!
(q^{x/\vartheta}\varOmega_{r-1};q)_\infty & := & \prod_{n_1=0}^{\infty}
(1-\varOmega_{r-1}q^{n_1+x/\vartheta})
\nonumber \\
& = &
\cR(s = (x/\vartheta + n_2\zeta_2/\vartheta +\cdots +n_r\zeta_r/\vartheta)(1-i\varrho(\vartheta))).
\label{G1}
\end{eqnarray}

Therefore the infinite products can be factorized as
\begin{eqnarray}
&& \prod_{n_2=0}^\infty\prod_{n_3=0}^\infty\cdots\prod_{n_r=0}^\infty\prod_{n_1=0}^\infty
(1-\varOmega_{r-1}q^{n_1+x/\vartheta})
\nonumber \\
 =\!\! && \prod_{n_2=0}^\infty\prod_{n_3=0}^\infty\cdots\prod_{n_r=0}^\infty
\cR(s = (x/\vartheta+n_2\zeta_2/\vartheta+n_3\zeta_3/\vartheta+\cdots +n_r\zeta_r/\vartheta)(1-i\varrho(\vartheta))).
\label{PR}
\end{eqnarray}
Using Eq. (\ref{A}) we have
\begin{eqnarray}
\fA_r(x\vert \epsilon) & = & \prod_{n_2=0,\ldots,n_r=0}^\infty\prod_{n_1=0}^\infty
(1- q^{x/\vartheta +n_1+n_2\zeta_2/\vartheta +\cdots +n_r\zeta_r/\vartheta})
\nonumber \\
& = &
\prod_{n_2=0,\ldots,n_r=0}^\infty\cR(s = (x/\vartheta+n_2\zeta_2/\vartheta+\ldots +n_r\zeta_r/\vartheta)(1-i\varrho(\vartheta))).
\label{A11}
\end{eqnarray}

\subsection{Symmetry and modular properties}
\label{Symmetry}

The next step of the iterative loop becomes the Jackson (convergent) double infinite product
$(q^{x/\vartheta};q,t)_\infty$ \cite{Jackson}, where $q=e^{2\pi i\vartheta},\, t=e^{2\pi i\tau}$,
\begin{eqnarray}
(q^{x/\vartheta};q,t)_\infty & = &
\prod_{n_1=0}^\infty\prod_{n_2=0}^\infty(1-q^{x/\vartheta +n_1+(\tau/\vartheta)n_2})
\nonumber \\
&=&
\prod_{n_2 = 0}^\infty \cR(s = (x/\vartheta + (\tau/\vartheta)n_2)(1-i\varrho(\vartheta))).
\label{n12}
\end{eqnarray}

For the product (\ref{n12}) two first order $q$- and $t$-equations
take the forms (see also \cite{Spiridonov})
\begin{eqnarray}
\frac{(q^{x/\vartheta}; q,t)_\infty}
{(qq^{x/\vartheta};q,t)_\infty} & = &
(q^{x/\vartheta};t)_\infty= \cR(s=x/\vartheta(1-i\varrho(\vartheta))+1-\tau/\vartheta),
\label{q1}
\\
\frac{(q^{x/\vartheta}; q,t)_\infty}
{(tq^{x/\vartheta}; q,t)_\infty} & = &
(q^{v/\vartheta}; q)_\infty = \cR(s=x/\vartheta(1-i\varrho(\vartheta))+1).
\label{q2}
\end{eqnarray}

Symmetry properties of Jackson double infinite product $(q^{x/\vartheta};q,t)$ analogous to
(modular) properties of the standard elliptic gamma functions.
For $z\in {\mathbb C}^\ast$ the order one $\Gamma_1$ and double (i.e., the order two) $\Gamma_2$
standard elliptic gamma functions have the forms
\begin{eqnarray}
\Gamma_1 (z;q,t) & = & \prod_{n_1,n_2 = 0}^\infty\left(\frac{1-z^{-1}q^{n_1+1}t^{n_2+1}}
{1- zq^{n_1}t^{n_2}}\right),
\nonumber \\
\Gamma_2(z; q,t,v) & = &
\!\! \prod_{n_1,n_2,n_3 = 0}^\infty
(1- z^{-1}q^{n_1+1}t^{n_2+1}v^{n_3+1})(1- zq^{n_1}t^{n_2}v^{n_3}).
\end{eqnarray}
The double elliptic gamma function $\Gamma_2$ has the following interesting modular
properties \cite{GR}:
\begin{eqnarray}
\Gamma_2(z; a, b, c) & = & \Gamma_2(z/a; -1/a,b/a,c/a)\cdot\Gamma_2(z/b; a/b,-1/b,c/b)
\cdot\Gamma_2(z/c; a/c,b/c,-1/c)
\nonumber \\
&\times &
\!\!{\rm exp}\left(\frac{i\pi}{12}B_{44}(z; a,b,c)\right)\,,
\end{eqnarray}
where $B_{44}$ is given by
\begin{equation}
B_{44}(z; a,b,c) = \lim_{\stackrel{x\rightarrow 0}{}}\frac{d^4}{dx^4}\frac{x^4e^{zx}}
{(e^{ax}-1)(e^{bx}-1)(e^{cx}-1)}.
\end{equation}

\section{Factorization for the partition function of the squashed sphere}
\label{Squashed}

In this section we describe an interesting factorization property and holomorphic blocks for the partition
functions of the squashed 3-sphere.
\footnote{A simple example to construct supersymmetry-preserving geometries which are topologically
the ellipsoid and which is
a deformation of the three-sphere, preserving a $U(1)\times U(1)$ isometry, can be parametrised as \cite{Pasquetti}:
$b^2\vert z_1\vert + (1/b^2)\vert z_2\vert^2 = 1,\, z_1,\,z_2 \in {\mathbb C}$, where $b$ is the squashing parameter.}

For the squashed sphere partition function it was observed in \cite{Pasquetti} that the partition function can be
written as:
\begin{equation}
 Z_{S_b^3}(m_a) = \sum_\alpha B_\alpha(x_a;q){\tilde B}_\alpha({\tilde x}_a; {\tilde q}).
 \label{Z}
 \end{equation}
In Eq. (\ref{Z}) the index $\alpha$ labels vacua of the mass-deformed theory, and $B_\alpha$ (respectively,
${\tilde B}_\alpha$) are certain holomorphic functions of $q=e^{2\pi ib^2}$ (respectively, ${\tilde q}=e^{2\pi ib^{-2}}$).
$x_a = e^{2\pi bm_a}$ and ${\tilde x}_a =e^{2\pi b^{-1}m_a}$.

\begin{remark}
Similar factorization was conjectured for the supersymmetric index \cite{Dimofte}, has been
described in a unified framework in \cite{Beem}. Similarly factorize was also shown for the lens space partition function
\cite{Imamura,Nieri} and also for the topological twisted index \cite{Nieri}.

In all mentioned cases, the partition functions of a given theory
on any of these manifolds are built out of the same objects $B_\alpha$, the so-called {\it holomorphic blocks}.
One of the interesting observation combined with the corresponding spaces of partition functions:
$S^3, S^3/{\mathbb Z}_p$ and $S^2\times S^1$ (see \cite{Willett} for detail); these spaces admit a Heegard
decomposition as a union of two solid tori,
$S^1\times D^2 \cite{Yoshida}$.
\end{remark}
In order to make the connection to the partition functions we can use the following relations \cite{Willett}:
${\tilde q} = g\cdot q,\,\, {\tilde x} = g\cdot x$, and $g$ implements the action of the diffeomorphism, acting as:
$
g = \left(\begin{matrix}
a\,\,\,  b \cr
c \,\,\, d
\end{matrix}\right) \in SL(2, {\mathbb Z})\Longrightarrow q= e^{2\pi i\vartheta} \rightarrow {\tilde q} =
e^{\pm 2\pi i\varphi(a, b\vert c, d;\, \vartheta)}$,
where $\varphi(a, b\vert c, d;\, \vartheta) = \frac{a\vartheta+b}{c\vartheta + d},\,\, x= e^{2\pi i\mu} \rightarrow {\tilde x} =
e^{\pm 2\pi i\varphi(0,\mu\vert c, d;\,\vartheta)}$.
In the case of a free chiral multiplet charged under a $U(1)$ flavor symmetry, a single block is  given by \cite{Willett}:
\begin{equation}
 B_\triangle(x; q) = (qx^{-1}; q)_\infty  =  \prod_{n=0}^\infty(1-q^{n+1}x^{-1})\stackrel{Eq.\, (\ref{R1})}{=\!=\!=\!=}
 \cR(s=(1-\mu/\vartheta)(1-i\varrho(\vartheta))),\,\,\,\, \vert q\vert< 1.
\end{equation}
At the same time
\begin{equation}
B_\triangle(x; q) = (qx^{-1}; q)_\infty  = \prod_{n=0}^\infty(1-q^{-n}x^{-1})^{-1}
= \cR(s= -\mu/\vartheta(1-i\varrho(\vartheta)+2)^{-1}, \,\,\,\,\vert q\vert> 1,
\end{equation}
where symbol $\triangle$ is connected to the Cartan generator of the $SU(2)$ R-symmetry factor.

The case of the S-fusing gives the $S_b^3$ partition function (see Eq. (\ref{Z})). Simplify calculations for this case
and taking $c=0$ $(ad=1)$ and $b^2$ to have positive imaginary part, so that $\vert q\vert, \vert {\tilde q}\vert^{-1}< 1$, we get:
\begin{equation}
B_\triangle(x; q)B_\triangle({\tilde x}; {\tilde q}) = \prod_{n=0}^\infty\frac{1-q^{n+1}x^{-1}}{1-{\tilde q}^n{\tilde x}^{-1}}
= \frac{\cR(s=(1-\mu/\vartheta)(1-i\varrho(\vartheta)))}{\cR(s= (\pm b/d\vartheta \mp \mu/d\vartheta)(1-i\varrho(\vartheta))+ 1 \mp a/d)}\,.
\end{equation}
This reproduce the $S_b^3$ partition function of a free chiral multiplet with the chosen contact terms.

Note that a similar result occurs for the other partition functions in a gauge field theory where are usually many blocks,
and a contour integral prescription for computing them was given in \cite{Beem}. In addition the blocks were derived directly
by localization theorem in \cite{Yoshida}.

\section*{Acknowledgments}

We are much grateful to Markku Oksanen for several remarks and
improvements in the work.
AAB would like to acknowledge the Conselho Nacional
de Desenvolvimento Cient\'{i}fico e Tecnol\'{o}gico (CNPq, Brazil) and
Coordenac\~{a}o de Aperfei\c{c}amento de Pessoal de N\'{i}vel Superior
(CAPES, Brazil) for financial support.

\end{document}